\documentclass[referee,traditabstract]{aa}
\usepackage{graphics}
\usepackage{dcolumn}
\usepackage[reqno]{amsmath}
\usepackage{amsfonts}
\usepackage{amssymb}
\usepackage{bm}
\usepackage{hyperref}
\usepackage{url}
\usepackage{subfigure}
\usepackage[sort]{natbib}
\usepackage{txfonts}
\usepackage{color}
\begin{document}
\newcommand{\changeref}[1]{\textbf{#1}}

\newcommand{\changeJ}[1]{\textcolor{blue}{#1}}
\newcommand{\changeJII}[1]{\textcolor{red}{#1}}
\newcommand{\TBB}{{{T_{\rm BB}}}}
\newcommand{\Te}{{{T_{\rm e}}}}
\newcommand{\Teq}{{{T^{\rm eq}_{\rm e}}}}
\newcommand{\Ti}{{{T_{\rm i}}}}
\newcommand{\nB}{{{n_{\rm B}}}}
\newcommand{\nS}{{{n_{\rm s}}}}
\newcommand{\Teff}{{{T_{\rm eff}}}}

\newcommand{\id}{{{\rm d}}}
\newcommand{\aR}{{{a_{\rm R}}}}
\newcommand{\neb}{{{n_{\rm eb}}}}
\newcommand{\kB}{{{k_{\rm B}}}}
\newcommand{\EB}{{{E_{\rm B}}}}
\newcommand{\zmin}{{{z_{\rm min}}}}
\newcommand{\zmax}{{{z_{\rm max}}}}
\newcommand{\YBEC}{{{Y_{\rm BEC}}}}
\newcommand{\YSZ}{{{Y_{\rm SZ}}}}
\newcommand{\rhob}{{{\rho_{\rm b}}}}
\newcommand{\Ne}{{{n_{\rm e}}}}
\newcommand{\sigT}{{{\sigma_{\rm T}}}}
\newcommand{\me}{{{m_{\rm e}}}}
\newcommand{\nBB}{{{n_{\rm BB}}}}

\newcommand{\kD}{{{{k_{\rm D}}}}}

\title{Does Bose-Einstein
  condensation of CMB photons cancel $\mu$ distortions created by dissipation of sound waves in the early Universe?}

\author{Rishi Khatri\inst{\ref{inst1}}
\and
Rashid A. Sunyaev\inst{\ref{inst1}{,}\ref{inst2}{,}\ref{inst3}}
\and
Jens Chluba\inst{\ref{inst4}}
}

\institute{ Max Planck Institut f\"{u}r Astrophysik, Karl-Schwarzschild-Str. 1,
  85741, Garching, Germany \email{khatri@mpa-garching.mpg.de}\label{inst1}
\and
 Space Research Institute, Russian Academy of Sciences, Profsoyuznaya
 84/32, 117997 Moscow, Russia \label{inst2}
\and
Institute for Advanced Study, Einstein Drive, Princeton, New Jersey 08540, USA\label{inst3}
\and
Canadian Institute for Theoretical Astrophysics, 60 St George Street, Toronto, ON M5S 3H8, Canada\label{inst4}}
\date{\today}

\abstract
{The difference in the adiabatic indices of photons and non-relativistic
  baryonic matter in the  early Universe causes the  electron temperature
  to be slightly lower than the radiation temperature. Comptonization of photons with colder electrons results in the
  transfer of energy from photons to electrons and ions owing to the  recoil
  effect {(spontaneous and induced)}.  Thermalization of
  photons with a colder plasma results in
 the accumulation of photons  in the Rayleigh-Jeans tail, aided by stimulated recoil,  while the higher frequency
 spectrum tries to approach Planck spectrum at the electron temperature $T_{\gamma}^{\rm final}=\Te<T_{\gamma}^{\rm initial}$; i.e.,  Bose-Einstein
  condensation of photons occurs. We find new  solutions of the Kompaneets
  equation describing this effect.  No actual condensate is, in reality,  possible since
  the process is very slow and  photons
  drifting to low frequencies are efficiently absorbed by bremsstrahlung and double
  Compton processes. {The spectral distortions created by Bose-Einstein
  condensation of photons are within an order of  magnitude (for the present
  range of allowed cosmological parameters), with exactly the same
  spectrum but opposite in sign, of
  those created by diffusion damping of the acoustic waves on small
  scales corresponding to comoving wavenumbers $45<  k< 10^4\,
  \rm{Mpc}^{-1}$.} The initial perturbations on these scales are completely
unobservable today due to their  being erased completely by Silk
damping. There is  {partial cancellation} of these two  distortions, leading to suppression of
$\mu$ distortions expected in the standard model of cosmology.   The net distortion  depends
 on the scalar power index $\nS$ and its running  $\rm{d} \nS/\rm{d}\ln k$.
 and {may vanish for special values of parameters, for example,
   for a running spectrum with, $\nS=1,\rm{d} \nS/\rm{d}\ln k=-0.038$.} We
arrive at an intriguing conclusion: even a null result, non-detection of
$\mu$-type distortion at a sensitivity of $10^{-9}$, gives a quantitative
measure of the primordial small-scale power spectrum.}

\maketitle
\section{Introduction}
In the early Universe  we have two strongly interacting fluids: cosmic
microwave background (CMB) radiation and
plasma (=ions+electrons).   Electrons and CMB are strongly
coupled due to the {Compton} process with the rate of energy transfer from
CMB to electrons given by ${R_{\rm compton}}\sim 1.4\times 10^{-20}(1+z)^4 \rm{s}^{-1}$. This
is much faster than the expansion rate , $H(z)\sim 2.1\times
10^{-20}(1+z)^2\rm{s}^{-1}$ \footnote{We give the formula that is valid during radiation domination just
  for illustration. The Hubble rate remains much lower than the
  Comptonization rate at $z>200$ and  than the Coulomb rate at even
  lower redshifts.},
 and keeps $T_{\gamma}-\Te\ll T_{\gamma}$
at redshifts $z\gtrsim 200$, where $T_{\gamma}$ is the effective temperature of
radiation (to be defined below)
and $\Te$ is the electron temperature. Ions and electrons exchange energy
via Coulomb collisions at a rate ${R_{\rm coulomb}}\sim 3\times
10^{-9}(1+z)^{3/2}\rm{s}^{-1}$ \citep{ll95} keeping $\Te-\Ti \ll\Te$, where $\Ti $ is the
temperature of ions. Thus we have {$T_{\gamma}\approx \Te\approx \Ti $} to high precision
at $z>200$.  Non-relativistic matter has {an} adiabatic index of $5/3$, while the
adiabatic index of radiation is $4/3$. Adiabatic cooling due to the
expansion of the Universe thus makes the matter cool faster than the
radiation, while Comptonization tries to maintain matter and radiation at
the same temperature by transferring energy from radiation to matter. As a
result, the energy of radiation decreases, while the number of photons is
conserved (neglecting absorption of photons by bremsstrahlung and double
Compton). This effect was discussed by \citet{zks68} and \citet{peebles68}
and leads to a decrease in the temperature of plasma compared to the radiation
temperature at $z\lesssim 200$, when Comptonization becomes inefficient
because of 
 depletion of electrons from recombination. It also drives the 21-cm line
spin temperature to below the radiation temperature, raising the possibility of
21-cm 
absorption of CMB photons by neutral hydrogen. This transfer of energy from
CMB to matter happens at all redshifts, as long as there is significant free
electron number density,  and results in deviations in CMB spectrum from the
blackbody.
Recently, \citet{cs2011} have numerically studied these
deviations from the {blackbody} in the CMB
spectrum. 
Below we  demonstrate that the problem of spectral deviations of CMB
due to loss of energy to plasma, in the case of small spectral
deviations and small Compton parameter $y$, has a simple analytic
solution, {when photon production and destruction is neglected}. The resulting flow of photons towards  low frequencies as the
 spectrum tries to approach the Planck spectrum is in fact the
 Bose-Einstein condensation of photons.   The 
Bose-Einstein condensation of photons
unfortunately does not progress very far in reality, because low-frequency
photons are efficiently absorbed by bremsstrahlung and double Compton
processes, and the {Compton} process {freezes out} as the electron density
falls owing to the expansion of the Universe and recombination.

Early papers about energy release in the early Universe were concerned with
exotic sources such as annihilation of matter and antimatter, primordial
turbulence, decay of new unstable particles, unwinding of topological
defects like domain walls, and cosmic strings.  Experiments 
beginning with COBE FIRAS \citep{cobe} have been unable to find any
significant distortions in the CMB from blackbody. New proposals like
{\sc Pixie} \citep{pixie} are demonstrating that a tremendous increase in the
sensitivity is possible in the future experiments. {\sc Pixie} is proposed to be 1000
times more sensitive than COBE FIRAS. There is hope that this is still not the
last word, {and} even higher sensitivity might {become} possible in the future. Under
these circumstances we decided to check what minimum levels of deviations
from blackbody spectrum are expected in the standard cosmological model
\citep{wmap7}. In this paper we only consider  the spectral distortions
arising before the end of the dark ages and beginning of reionization. 
We 
demonstrate that, in the absence of decay and annihilation of new unknown
particles or any other new  physics beyond the standard model, there are
only three key reasons for significant global spectral distortions in the CMB.\\
1. Bose-Einstein condensation of CMB photons due to the difference in the
adiabatic indices of non-relativistic plasma and photons.\\
2. The energy release due to dissipation of sound waves and initial
perturbations due to Silk damping \citep{silk} and the second-order Doppler
effect due to non-zero peculiar velocity of electrons and baryons in the
CMB rest frame, both leading to the superposition of {blackbodies} {\citep{Zeldovich1972}} in the electron
rest frame.\\
{3. The cosmological recombination radiation from hydrogen and helium \citep{zks68, peebles68, Dubrovich1975}}{.}

{We  omit the distortions caused by the cosmological
  recombination process \citep[e.g., see][for more details and references
  therein]{Chluba2006, Sunyaev2009} in the discussion below.  They
  have   narrow features and different continuum spectrum  and can be
  distinguished from $y$ and $\mu$ distortions from the first two
  mechanisms. Cosmological {recombination} radiation 
  can, however, become comparable to or larger than those discussed here \citep{cs2011},
  if the distortions from the first and second processes
  partially cancel.}
We also note  that for any experiment with finite beam size there will be a
mixing of {blackbodies} in the beam owing to the temperature
fluctuations on the last scattering surface on scales smaller than the beam
size, leading to inevitable $y$ distortions of magnitude $\sim (\Delta
T/T)^2\sim 10^{-9}-10^{-10}$ {\citep{Chluba2004}}.

In this paper we find new solutions for the Kompaneets equation describing
the CMB spectral distortions arising from  {process 1} described above.
A second mechanism results in a {$y$-distortion} and heating of the
  electrons at low redshifts. At high redshifts   Comptonization of CMB on the hotter electrons
converts the {$y$-type} distortion  to a $\mu$-type distortion (\citet{sz1970};\citet{is1975}). We
 demonstrate below that {these } two {sources} of distortions work against
each {other, individually resulting} in spectral distortions {with} opposite signs but {the} same spectral shape in any part of the CMB spectrum.

The cooling of electrons and the corresponding spectral distortions are
easy to calculate and only depend  on the standard cosmological parameters,
such as baryon to photon number density $\nB/n_{\gamma}$ and helium fraction, which decide the
amount of energy losses by the CMB, the Hubble constant
($H_0$), and densities of constituents of the Universe, which in turn {determine} the
expansion rate and thus the efficiency of {Comptonization}.
{On the other hand,} the energy released by dissipation of sound waves
crucially depends on the power in small-scale fluctuations, and thus the
spectral index ($\nS $) in the standard cosmological model, in addition to
the other well-measured parameters of the standard cosmological model,
\citep{wmap7}. It is interesting to note that for a spectrum with constant $\nS$ the energy
release due to dissipation exceeds the energy losses to adiabatic cooling
of baryons, and there is net heating of electrons. For a primordial spectrum
with a running spectral index, the role of {Comptonization}
with colder electrons and Bose-Einstein condensation can become dominant, with
the spectral distortions changing sign, and there can be net cooling of electrons
and a corresponding decrease of CMB entropy per baryon (specific
entropy). Additional increase in entropy during the recombination epoch due to
superposition of {blackbodies}, free streaming as well as Silk damping and
the second-order Doppler effect only produce  $y$-type 
distortions and can be distinguished from the $\mu$ distortions created in
the earlier epoch.

\section{Thermodynamic equilibrium in the early Universe}
We can obtain the energy losses due to adiabatic cooling of baryons  under the assumption
that Compton scattering, bremsstrahlung, and double Compton scattering can
maintain full thermodynamic equilibrium between the
electrons and photons. This is true to high accuracy at $z>10^6$ while at
lower {redshifts, although} bremsstrahlung and double Compton scattering
cannot {help in restoring a} Planck spectrum at all frequencies, Comptonization keeps {the} electron temperature equal to the effective radiation temperature ({cf.} Eq.~\eqref{te}) to high
{precision} until $z\sim 200$. In the thermodynamic equilibrium, entropy is
conserved, and we can  {use this}  to calculate the energy losses
due to the expansion of the Universe without referring to {Comptonization},
since it is not really important which physical process is responsible for
the equilibrium.

We start with the thermodynamic relation,
\begin{align}
T_{\gamma}\id S=\id U+P \id V-M \id N,\label{thermeq}
\end{align}
where $S$ is the entropy, $U$ total thermal energy, $P$ pressure, $V$
volume, $M$ chemical potential, and $N$ the number of particles. $T_{\gamma}$
is the common temperature of photons, ions, and electrons.
We can ignore the last term for ions and electrons since their number is
fixed after big bang nucleosynthesis ends\footnote{During recombination, the number of particles
  changes and the departure from thermodynamic equilibrium also becomes
  significant. The following calculations are therefore strictly valid only
before recombination.}. For photons as well we can ignore
the last term if we assume that their chemical potential is $0$, which is
true in full thermodynamic equilibrium.
With this assumption we can write the equation for total entropy per
baryon, $\sigma=s/\nB$, where $s$ is the entropy density and $\nB$ is the
baryon number {density:} 
\begin{align}
\id\sigma =\frac{\id(E/\nB)+P\id(1/\nB)}{T_{\gamma}}\label{dentropy}\\
E=\aR T_{\gamma}^4+\frac{3}{2}N\nB \kB T_{\gamma}\\
P=\frac{1}{3}\aR T_{\gamma}^4+N\nB \kB T_{\gamma}{.}
\end{align}
{Here} $\aR$ is the radiation constant and $N$ the number of non-relativistic particles per baryon, so that
$N\,\nB =\neb $.
We now integrate Eq.~\eqref{dentropy} to {obtain} 
\begin{align}
\sigma=\frac{4 \aR T_{\gamma}^3}{3\nB }+ N \kB \ln\left(\frac{T_{\gamma}^{3/2}}{\nB C}\right),
\end{align}
{where} $C$ is an arbitrary constant of integration, which is not important for our
calculation. The first term above is the contribution from photons, and it
is clear that in the absence of second term, $T_{\gamma}\propto (1+z)$. The
second term is the contribution of non-relativistic particles and causes
the temperature to drop slightly faster in order to conserve {the} entropy per
baryon. We can write $T_{\gamma}=T(1+t)$, where
$T=\Ti (1+z)/(1+z_{i})$ is the background temperature proportional to
$1+z$, and $t\ll 1$ is the fractional deviation from this law. We can take the initial
deviation $t(z_i)=0$ at initial redshift $z_i$ and then calculate the
subsequent energy losses to adiabatic cooling of baryons  at later
redshifts. Thus we can write {$\sigma$, correct to leading order in $t$,} as 
\begin{align}
\sigma=\frac{4 \aR T^3}{3\nB }+ \frac{4 \aR T^3 (3t)}{3\nB }+N\kB
\ln\left(\frac{T^{3/2}}{\nB C}\right) + \frac{3N\kB t}{2}\label{entropy}{.}
\end{align}
Equating the above to {the} initial entropy $\sigma_i=\sigma(z_i)$ with {$t\equiv 0,T\equiv \Ti$}, we get {upon} solving for $t$ (and ignoring the last term in Eq.~\eqref{entropy} compared to
the second term)
\begin{align}
t&=\frac{3N\nB \kB }{8 \aR T^3}\ln\left(\frac{1+z}{1+z_i}\right)\nonumber\\
&=\frac{\EB}{4E_{\gamma}}\ln\left(\frac{1+z}{1+z_i}\right)\nonumber\\
&=-\frac{5.9\times 10^{-10}}{4}\ln\left(\frac{1+z_i}{1+z}\right)\label{thermres}
\end{align}
{with $\Delta E/E=4t$, since
$E_{\gamma}\propto T^4$, and $\EB=\frac{3}{2}\kB \neb  \Te$}. we note that $t$ is negative since $z_i>z$ and there
is net energy loss. {$\ln (z_{\rm max}/z_{\rm min})\sim 10$} 
for the redshift range of interest, $200<z<2\times 10^6$, where {a mixture of $\mu$ and $y$-type} distortions are produced.
{This result for the total energy extraction is consistent with the
  estimate obtained by \citet{Chluba2005} and \citet{cs2011}.}

To estimate the magnitude of total distortions, $\mu$ and $y$-type,  we
take  $\zmax=z_i=2\times 10^6$, since at greater redshifts spectral distortions are
rapidly destroyed by bremsstrahlung and double Compton processes and
{Comptonization}. {The latter} redistributes photons over the whole spectrum. {For $\zmin=200$ we get $\Delta E/E=5.4\times 10^{-9}$.}
This transfer of energy from radiation to baryons due to a difference in
their adiabatic indices results in an
inevitable distortion and  Bose-Einstein condensation of photons. We
 show below that the distortion has a magnitude  of  {$
\YBEC\sim(1/4)\,\Delta E/E\sim  10^{-9}$}. We  define
parameter $\YBEC$ and prove this result below.
The corresponding deviation of the electron temperature from the equilibrium temperature of electrons in the radiation field is
described by Eq.\eqref{te} and is  $(T_{\gamma}-\Te)/T_{\gamma}\approx 10^{-12}$
at $z=10^6$  growing to $\sim 0.01$ at $z=500$.  

{In the Universe} there are also inevitable processes like
{the} dissipation of acoustic waves and SZ effect from reionization, which would
lead to a normal SZ effect of similar or greater magnitude. Due to the
additivity of small spectral distortions,  these effects will cancel the
$\YBEC$ distortions, and this diminishes the hope that this 
Bose-Einstein condensation  will
ever be observed. Nevertheless it is important to stress that in the
standard cosmology there was an epoch $10^6>z>1000$ when 
Bose-Einstein 
condensation of CMB photons due to the difference in the adiabatic indices
of matter and radiation was able to create a peculiar deviation of CMB spectrum from
the blackbody.

\section{Alternative direct calculation of energy losses in CMB}
We would like to  remind the reader that adiabatic cooling of baryons leads to the cooling of
radiation, but Comptonization itself conserves the number of
photons. We can write the equation for the evolution of average  thermal energy density
{$\EB$} in
baryons before recombination as
\begin{align}
\frac{\id E_{B}}{\id z}&\equiv\frac{3}{2}\kB \Te
\frac{\id\neb }{\id z}+\frac{3}{2}\kB \neb  \frac{\id\Te}{\id z}.
\end{align}
The derivative in the first term on the right-hand {side, $\id\neb /{\id z}=3\neb /(1+z)$,} is just the decrease in number
density of  particles (ions and electrons) with redshift $z$ due to {the
  Hubble} expansion, with $\neb =\rhob/\mu_{\rm mol}$, where $\rhob$ is the baryon mass density and
$\mu_{\rm mol}$  the mean molecular weight. The change in temperature has
contributions from adiabatic cooling and also from energy gained from CMB by
Comptonization:
\begin{align}
\frac{\id\Te}{\id z}= \frac{2 \Te}{1+z} - \frac{2}{3 \kB  \neb }{S_{\rm Compton}}\label{tee}
\end{align}
where ${S_{\rm Compton}}=4\kB   E_{\gamma}\Ne \sigT(T_{\gamma}-\Te)/ \me c
H(z)(1+z)$ is the energy transfer rate per unit volume from radiation to baryons by
Compton scattering, $\Ne $ is the number density of free electrons,
$\sigT$  the Thomson scattering cross section, $\kB $  the Boltzmann
constant, $E_{\gamma}$  the energy density of radiation, $c$  the speed
of light, {$\me$  the mass of electron}, and $H(z)$  the Hubble parameter.   That redshift decreases
with increasing time, so the terms with {``$+$"} sign are cooling terms and
terms with {``$-$"} sign are heating terms.
The change {in} photon energy density can therefore be written as
\begin{align}
\frac{\id E_{\gamma}}{\id z}&=\left(\frac{\id E_{\gamma}}{\id z}\right)^{{\rm adiabatic}}+\left(\frac{\id E_{\gamma}}{\id z}\right)^{{\rm Compton}}\nonumber\\
&=\frac{4E_{\gamma}}{1+z}+{S_{\rm Compton}},
\end{align}
where the first term is just the adiabatic cooling due to the expansion of
the Universe.
We can estimate the energy transfer,
{$S_{\rm Compton}$}, by noting that the baryon temperature $\Te=T_{\gamma}\propto 1+z$ to a
high accuracy {until $z\sim200$}, where {$T_{\gamma}=2.725(1+z)\,\rm K$} is the CMB
temperature.  Using this to evaluate the total derivative of
electron temperature $\id \Te/\id z$ on the left-hand side of Eq.~\eqref{tee} and
the first term on the right-hand side, we
get 
\begin{align}
{S_{\rm Compton}}
&=\frac{3}{2}\kB \neb \frac{2T_{\gamma}}{1+z}-\frac{3}{2}\kB \neb 
\frac{\id T_{\gamma}}{\id z}\nonumber\\
&={\frac{\EB}{1+z}.}
\end{align}
Thus we have fractional rate at which  energy is lost by radiation to baryons
\begin{align}
\frac{(\id E_{\gamma}/\id z)^{{\rm Compton}}}{E_{\gamma}}=\frac{{S_{\rm Compton}}}{E_{\gamma}}=\frac{\EB/E_{\gamma}}{(1+z)}=\frac{5.9\times 10^{-10}}{1+z}.\label{eeq}
\end{align}

We can now calculate the total \emph{fractional} energy losses of radiation
that
contribute to the spectral distortions   between redshifts
$\zmin$ and $\zmax$,
\begin{align}
\frac{\Delta E}{E}&=\int_{\zmax}^{\zmin}\frac{\id E_{\gamma}}{E_{\gamma}}\nonumber\\
&=\int_{\zmax}^{\zmin}\frac{\id z}{E_{\gamma}}\left(\frac{\id E_{\gamma}}{\id z}\right)^{{\rm Compton}}\nonumber\\
&=-5.9\times 10^{-10}\ln\left(\frac{1+\zmax}{1+\zmin}\right){.}\label{energy}
\end{align}
We have integrated $\id E/E$, since immediate distortions are proportional to
$\Delta E/E$, and the distortions can be added linearly if they are
small.  This is the same result as Eq.~\eqref{thermres}.

\section{Kompaneets equation}
The interaction of radiation with electrons through Compton scattering or
Comptonization is described by {the} Kompaneets equation \citep{k1956} in the Fokker-Planck
approximation, when the energy transfer in each scattering is small
compared to temperature, {and the incoming photon distribution is wide compared to the width of the scattering kernel:}
\begin{align}
\frac{\partial n}{\partial y}=\frac{1}{x^2}\frac{\partial }{\partial
  x}x^4\left(n+n^2 +\frac{\Te}{\TBB}\frac{\partial n}{\partial x}\right){,}\label{komp}
\end{align}
{where we have defined {the Compton parameter}}
\begin{align}
y(z,z_{\rm{max}})=-\int_{z_{\rm{max}}}^{z}\id z\frac{\kB \sigT}{\me
  c}\frac{\Ne \TBB}{H(1+z)}{,}\label{yz}
\end{align}
which is convenient to use instead of time or redshift. We start our
calculation at 
  the reference redshift $z_{\rm{max}}$.
One can furthermore introduce the Compton equilibrium electron temperature in a radiation field \citep{zl1970, ls1971}
\begin{align}
\frac{\Teq}{\TBB}=\frac{\int n(1+n)x^4 \id x}{4\int n x^3 \id x}.\label{te}
\end{align}
Here $\TBB$ is a reference temperature which is equal to the radiation
blackbody temperature if the initial radiation field is a Planck spectrum, and 
$x=h\nu/\kB \TBB$ is the dimensionless frequency with $\nu$  the frequency of photons.
The equations written in this way factor out the expansion of the Universe
and are applicable to cosmology, as well as non-expanding astrophysical
systems. {This equation and the Kompaneets equation
  form a coupled system to be solved  simultaneously.}

The three terms in the inner brackets {of} Eq.~\eqref{komp} describe recoil
($n$), induced recoil ($n^2$), and {the} Doppler effect of the thermal motion of
electrons ($\Te/\TBB\partial n/\partial x$).
For $y\ll 1$ and an
initial blackbody spectrum with temperature {$\TBB$, $\nBB=1/(e^{x}-1)$,}
a simple first-order correction to the blackbody spectrum  can be
found \citep{zs1969}. By approximating\footnote{{For a Bose-Einstein distribution with constant chemical potential this expression is an identity.}}
$n+n^2\approx -\partial n /\partial x$ in Eq.~\eqref{komp} we get  
\begin{align}
\frac{\partial n(x,y)}{\Delta \partial y}=&\frac{1}{x^2}\frac{\partial}{\partial
  x}\left[x^4\frac{\partial n(x,y)}{\partial
      x}\right]\label{szeq},
\end{align}
where $\Delta\equiv \Te/\TBB-1$. 
This particular form of equation emphasizes that the Doppler term dominates over recoil and  gives rise to the spectral distortion by boosting low-energy photons to higher frequency. 
The above equation can be reduced to {the} diffusion or heat  equation
\citep{zs1969} by {changing variables},  for which {the} Green's
function is well known.

For small distortions, the solution takes a particularly simple form and can be found by
substituting $\nBB$ on the right-hand side of Eq.~\eqref{komp} or Eq.~\eqref{szeq}{:}
\begin{align}
\Delta n \equiv n(x,\YSZ )-\nBB(x) = \YSZ \frac{xe^x}{\left(e^x-1\right)^2}\left[x\frac{e^x+1}{e^x-1}-4\right]{,}\label{szap}
\end{align}
where we have defined $\YSZ =\int_0^y \Delta \id y$. The SZ parameter is
related to fractional energy release by the formula $\YSZ =(1/4) \Delta
E_{\gamma}/E_{\gamma}$. This can be obtained by  integrating the
above spectrum over all frequencies.

\section{New solution of Kompaneets equation for Comptonization of photons with colder plasma}
 The Kompaneets equation describes, to lowest order,  the
competition between the Doppler boosting of low-energy photons to high
energies and the down scatter of high-energy photons to low energies by recoil
and stimulated recoil.  The initial radiation spectrum is
blackbody with temperature $\TBB$. If the electrons are also at the same
temperature, $\Te=\TBB$, then these two effects cancel each other out exactly.
If $\Te>\TBB$, the Doppler boosting becomes stronger than recoil and we
have the normal SZ effect {in the limit of small $y$ parameter}. If $\Te< \TBB$ then Doppler boosting is weaker
and we  have a net movement of photons from high to low energies, which is
the complete  opposite
of the SZ effect. But since at $\Te=\TBB$ the two effects exactly balance
each other out, at linear order the spectral distortions for $\Te<\TBB$ would
be exactly the same as  for $\Te>\TBB$, {just} with the opposite sign. 

For $\Te<\TBB$, we can write the approximate Kompaneets equation by
approximating   $\partial n/ \partial x\approx -(n+n^2)$  in
Eq.~\eqref{komp},
\begin{align}
\frac{\partial n}{\partial y}=(1-\frac{\Te}{\TBB})\frac{1}{x^2}\frac{\partial }{\partial
  x}x^4\left(n+n^2 \right).\label{bekomp}
\end{align}

Analytic solutions can be obtained by the standard method of
characteristics for the recoil ($n$)  and induced recoil
($n^2$)  terms individually, and they consist of photons moving down the
frequency axes at a  speed proportional to $x^2$ for the recoil term
\citep{arons,is72} and at a speed
proportional to the  $x^2n$ for the induced recoil term (\citet{zl1969};\citet{s1971}).
To get the linear solution, we can just substitute $n(x)=\nBB(x)$ on the
righthand side of Eq.~\eqref{bekomp} and define the parameter for the
amplitude of distortion {in} the case $\Te<\TBB$ as
$\YBEC=\int_0^y(1-\Te/\TBB)\id y$,  the result is  Eq.~\eqref{szap} with
$\YSZ =-\YBEC$. We should emphasize that the linear {solution, Eq.~\eqref{szap},}
can be arrived at in a straightforward way by just substituting $n(x)=\nBB(x)$
on the righthand side of the full Kompaneets equation, Eq.~\eqref{komp},  without any restriction on
$\Te$. However the dominant physical effects when $\Te>\TBB$ (Doppler) and
when $\Te<\TBB$ (recoil and induced recoil) are completely different and  that
in equilibrium they  balance each other out exactly  gives us the same mathematical form
of the linear solution in both cases.

\begin{figure}
\includegraphics{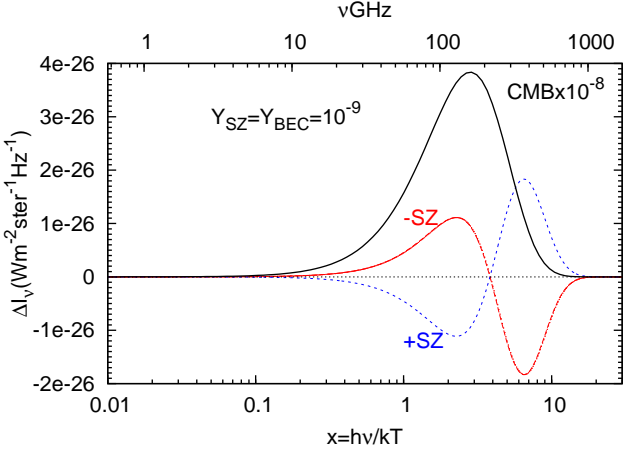}
\caption{\label{sz}Difference in intensity from the
blackbody radiation for normal SZ effect with $\YSZ =10^{-9}$ and negative
SZ effect with $\YBEC=10^{-9}$.}
\end{figure}

\begin{figure}
\includegraphics{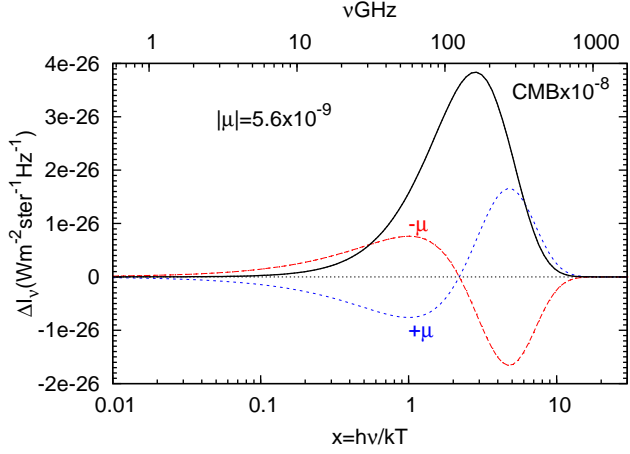}
\caption{\label{bose}Difference in intensity from the
blackbody radiation for  $\mu$ distortion defined by the equation
with $n(\nu)=1/(e^{h\nu/\kB T_{\mu}+\mu}-1)$. 
$T_{\mu}$ is the temperature
to which the spectrum relaxes for an initial energy addition/loss of $\Delta
E/E$  with $\Delta T_{\mu}/T=0.64\Delta E/E=\pm 2.5\times 10^{-9}$. $\Delta E/E=\pm 4\times 10^{-9}$
is the energy addition/loss that gives rise to SZ distortion of $10^{-9}$.
The dimensionless chemical potential $\mu$ is given by $\mu=2.2\Delta T/T=\pm 5.6\times 10^{-9}$.}
\end{figure}

In the Rayleigh-Jeans part of the spectrum we now have an increase in the
brightness temperature of $\Delta T/T=+2\YBEC$, which is independent of
frequency and thus maintains the Rayleigh-Jeans shape of the spectrum. In the Wien part we
have a decrease in intensity $\Delta I/I=\Delta n/n{\approx} -x^2
\YBEC$. Figure \ref{sz} shows the difference in intensity from the
blackbody radiation for normal SZ effect with $\YSZ =10^{-9}$ and
the negative SZ effect with $\YBEC=10^{-9}$. Figure \ref{bose} shows  the spectrum
that would be achieved at a
high value of the Compton parameter $y$. It is a Bose-Einstein spectrum
with  the dimensionless chemical potential\footnote{ This
  definition has a sign difference with respect to the usual definition of the
  chemical potential in thermodynamics \citep{llstats} used in Eq. \eqref{thermeq}.} $\mu$ defined by the equation
with $n(\nu)=1/(e^{h\nu/\kB T_{\mu}+\mu}-1)$, where $T_{\mu}$ is the temperature
to which the spectrum relaxes for an initial energy addition/loss of $\Delta
E/E$  with $\Delta T_{\mu}/T=0.64\Delta E/E=\pm 2.5\times 10^{-9}$
\citep{is1975b}, and $\Delta E/E=\pm 4\times 10^{-9}$
is the energy addition/loss that gives rise to SZ distortion of $10^{-9}$.
The chemical potential $\mu$ is given by\footnote{These relationships can be easily derived using photon
  number and energy conservation and requiring that the final spectrum have
  the equilibrium Bose-Einstein distribution.} $\mu=2.2\Delta T/T=\pm 5.6\times
10^{-9}$. This is the
spectrum that an initial spectrum with $\YBEC,\YSZ =10^{-9}$ will approach at
high $y$ at $x{\gg}\mu\sim10^{-9}$.  The frequency at which the distortion
crosses zero is at $x=2.19$ compared to $x=3.83$ for the SZ distortion in
Fig. \ref{sz}.

\begin{figure}
\includegraphics{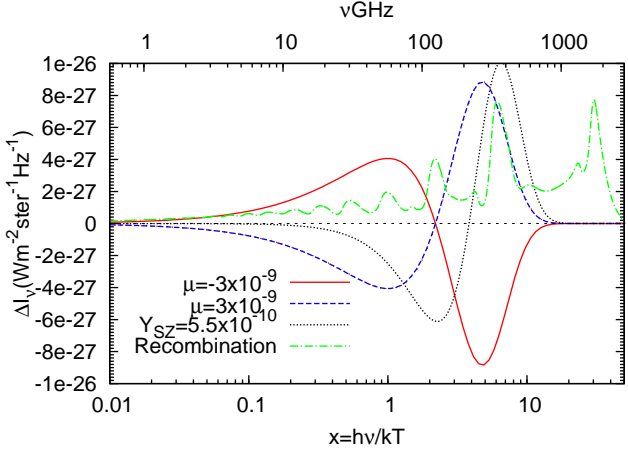}
\caption{\label{recom}Comparison of positive and negative  $\mu-$type distortions expected in the
  early Universe (to be calculated in  later sections) with the
  $y$-type distortions before reionization and 
 the cosmological recombination spectrum \citep[taken from][]{Chluba2006, Jose2008} for illustration.     }
\end{figure}

The $y$-type and $\mu-$type distortions expected in the early Universe (
calculated in the later sections) are
compared with the cosmological recombination spectrum \citep{rcs06} in Fig. \ref{recom}. Clearly $\mu$
type distortions have a different spectral shape than the recombination radiation (both from hydrogen \citep{Chluba2006} and helium \citep{Jose2008}) and $y$-type distortions and can be distinguished from the
last two. This is very important because  the information in the $\mu$ type
distortions  about the early Universe physics can be extracted. On the
other hand, the  $y$-type distortions
from the early Universe get swamped by the much larger $y$-type distortions
from the low redshifts and the two contributions are difficult to separate.
The $\mu$ distortions expected from the early Universe also have higher magnitude than the
recombination spectrum in the Rayleigh-Jeans part of the spectrum but has
no quasi-periodic structure like the cosmological recombination radiation.

\begin{figure}
\includegraphics{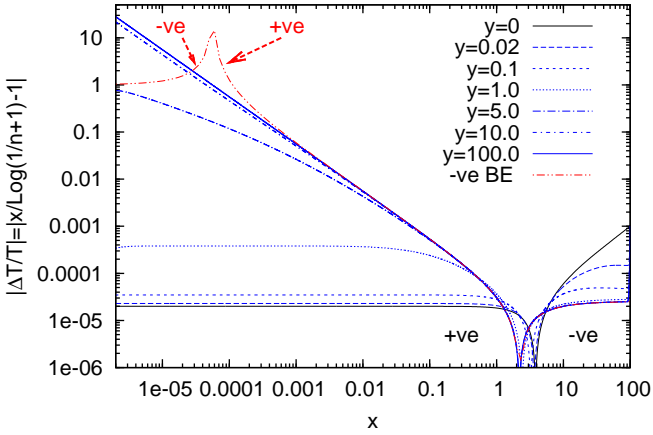}
\caption{\label{szfig}Evolution of initial spectrum at $y=0$ given
by Eq.~\eqref{szap} with $-\YSZ =\YBEC=10^{-5}$. $|\Delta T/T|\equiv |\Teff /T-1|$ is
plotted, where $\Teff $ is the  temperature of a blackbody
  spectrum with the corresponding intensity at a particular frequency.   Bose-Einstein spectrum defined by
$n(\nu)=1/(e^{h\nu/\kB \Te+\mu}-1)$ with negative $\mu$ is labeled {`-ve BE'},
red dashed, double-dotted line in
the figure.   The distortions are positive in the Rayleigh-Jeans
region (labeled '+ve') on the left side and negative in the Wien region
(labeled '-ve') on the right side
of the figure.}
\end{figure}

This flow of photons toward lower frequencies as the spectrum tries to approach
the Planck spectrum due to recoil, and induced recoil is  Bose-Einstein condensation of photons \citep{is1975b,llstats}.
We show the evolution of {the} spectrum (solution of the coupled
system of Eqs. \eqref{komp}
and \eqref{te}) starting with the initial {distortion} given
by Eq.~\eqref{szap} with $-\YSZ =\YBEC=10^{-5}$ in {Fig. \ref{szfig}.
In
  the Rayleigh-Jeans region intensity $I_{\nu}\propto \Teff $ and
  therefore $\Delta T/T = \Delta I/I$.
The initial} evolution is
similar to that of a spectrum with positive $\YSZ $ with the {photon distribution}
approaching a Bose-Einstein spectrum defined by
$n(\nu)=1/(e^{h\nu/\kB \Te+\mu}-1)$ with negative $\mu$ (marked {`-ve BE'} in
the figure).  We should emphasize that there is no singularity in the actual
solutions of the Kompaneets equation plotted above. The singularity is just in
the mathematical formula, which correctly describes the spectrum at
high frequencies, $x\gg|\mu|$. The actual spectrum  deviates from the
Bose-Einstein spectrum near the singularity (positive everywhere  in
the Rayleigh-Jeans region) and can be described by a
chemical potential decreasing  in magnitude with decreasing frequency. The evolution at $y>1$ is therefore very
different from the positive $\YSZ $ case. For $x\gg |\mu|$ the spectrum is  the
Bose-Einstein spectrum at the electron temperature. For $x\ll |\mu|$ there is an
excess of photons compared to the Bose-Einstein spectrum which grows with time, a feature of
Bose-Einstein condensation. The fractional change in temperature $\Delta T/T$ takes a very simple form in the
Rayleigh-Jeans region ($x\ll 1$) for a Bose-Einstein spectrum and can be
understood as follows. It is given, in terms of frequency
referred to the electron temperature, $x_e\equiv h\nu/kT_e$, by 
\begin{align}
\frac{\Delta T}{T}=\frac{x_e}{x_e+\mu}-1=\frac{-\mu}{x_e+\mu}.
\end{align}
For $|\mu|\ll x_e\ll 1$ we have $\Delta T/T=-\mu/x_e$. Thus the fractional
temperature deviation is positive for negative $\mu$ and negative for the
positive $\mu$. For $x_e$ greater than and close to $|\mu|, \mu<0$ the
spectral distortions can become very large and exceed unity.
Since the high-frequency
spectrum is in equilibrium at the electron temperature by $y=10$, the
subsequent evolution is very slow, and  $|\mu|$ decreases while the electron
temperature increases slowly as the {photon distribution approaches a} Planck spectrum with the extra photons accumulating at low
frequencies; i.e., Bose-Einstein condensation happens. The low-frequency
spectrum approaches the stationary solution of Kompaneets equation with
only the induced recoil term, $n(x)\propto 1/x^2$, with a continuous flow of
photons towards $x=0$ \citep{s1971}. Then, $\Te/T-1$ just increases from $-2.558\times 10^{-5}$ to
$-2.555\times 10^{-5}$ in  going from $y=10$ to $y=100$ for the chosen energy losses of $\Delta
E/E=4\times 10^{-5}$. The effect discussed below in the real Universe is of
magnitude $\sim 10^{-9}$.

\section{  Bose-Einstein condensation of CMB photons in the  early Universe.}
\begin{figure}
\includegraphics{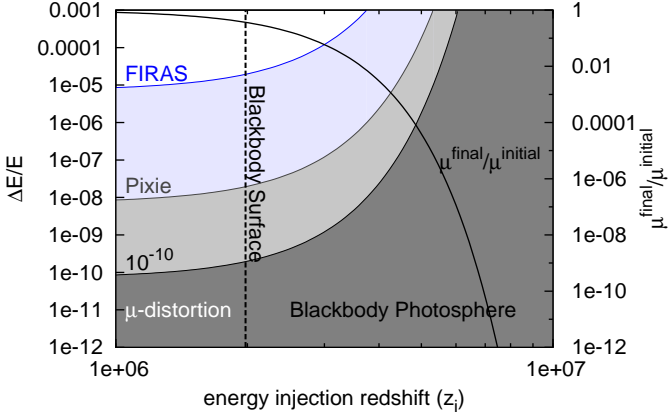}
\caption{\label{bbfig}Regions of the blackbody photosphere probed by
  different experiments, COBE FIRAS with sensitivity  $\mu=10^{-5}$, Pixie
  with sensitivity $\mu=10^{-8}$ and a hypothetical experiment with
  sensitivity of $\mu=10^{-10}$. Shaded regions below the curves are
  allowed and inaccessible to these experiments. We also plot the ratio of final $\mu$ distortion today  to the initial $\mu$ distortion  as a function of energy
  injection redshift. We define the blackbody surface as the energy
  injection redshift such that 
 $\mu^{\rm{final}}/\mu^{\rm{initial}}=1/e$, {$z_{\rm bb}=1.98\times
   10^6$}.}
\end{figure}
In the early Universe, before $z\sim 2\times 10^6$  double
Compton and {Comptonization} destroy any spectral distortions and
maintain the Planck spectrum of CMB. For small distortions due to single,
quasi-instantaneous episode of energy release, we can write the ratio of
final-to-initial $\mu$ as an exponential function of redshift defined by
the square root of the product of Comptonization and absorption rates
\citep{sz1970}. For a  double Compton process, this formula gives \citep{dd1982}
\begin{align}
G(z)=\frac{\mu^{\rm{final}}}{\mu^{\rm{initial}}}\approx e^{-(z_{\rm{i}}/z_{\rm{dc}})^{5/2}},
\end{align} 
 where $z_{\rm dc}\approx 1.98\times 10^6$ defines the ``surface of the
 blackbody photosphere". We call this function ${G}(z)$ the blackbody visibility function for spectral distortions. 
We call the region $z>z_{\rm{dc}}$ where the initial $\mu$ distortion can be reduced by a factor of more than  $e$ the ``blackbody photosphere". Thus inside the blackbody photosphere, a Planckian spectrum can be
  established efficiently. Figure \ref{bbfig} shows how the $\mu$ distortion at high
redshifts decreases due to the double Compton and Compton scattering.
The regions of the blackbody photosphere allowed by different experiments are
shaded.

\begin{figure}
\includegraphics{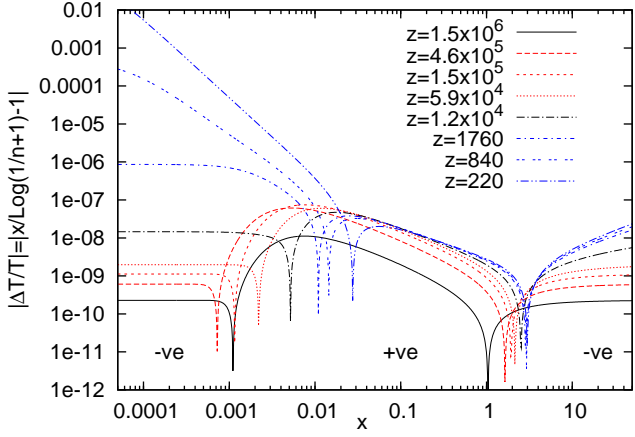}
\caption{\label{becfig}Evolution of spectral distortions in CMB in
  $\Lambda$CDM cosmology. The values of total Compton parameter $y=y(100,z)$  for different
  curves are $(z=1.5\times 10^{6},y=109)$, $(z=4.6\times 10^{5},y=10)$,
  $(z=1.5\times 10^{5},y=1)$, $(z=5.9\times 10^{4},y=0.16)$, $(z=1.2\times
  10^{4},y=0.005)$, $(z=1760,y=3\times 10^{-5})$, $(z=840,y=1.8\times
  10^{-8})$ and $(z=220,y=1.2\times 10^{-10})$.  Reference spectrum is the blackbody CMB spectrum
  at $z=10^7$. At low frequencies the spectrum is dominated by
  double Compton and bremsstrahlung which maintain Planck spectrum at the
  electron temperature. The zero point at low frequencies moves to the
  right at low redshifts because bremsstrahlung dominates over the Compton
  scattering at higher and higher frequencies and absorbs photons, although both are slower than the expansion rate.}
\end{figure}

At redshifts $10^5<z<2\times 10^6$, we have
the Compton parameter $y>1$, and the Bose-Einstein spectrum with a negative
chemical potential is established at $x\gtrsim 0.01$\footnote{{Strictly
  speaking, a Bose-Einstein spectrum is established at $y\sim few$ as can be
seen from Fig \ref{szfig} However at $y\sim 0.25 - 1$ the spectrum is already very
close to Bose-Einstein at high frequencies. We  use this to divide the
energy release into $\mu$-type and $y$-type estimates. Exact results are
presented elsewhere \citep{cks2012}}}. At lower
frequencies, bremsstrahlung and double Compton create {a} Planck spectrum
corresponding to the electron temperature. At $z<10^4$ the Compton
parameter $y<0.01$ and the distortions created can be described by the
linear solution of Eq.~\eqref{szap}. Thus the net  distortions created in the
early Universe owing to adiabatic cooling of baryons are {a} linear
superposition of distortions corresponding to different values of $y$, and a few
of them are
shown in {Fig.~\ref{becfig}}. 
We show the evolution of actual spectrum (Eq.~\ref{komp} and \ref{te})
including the effects of bremsstrahlung and double Compton processes
{\citep{is1975, is1975b, hs1993, bdd91, cs2011}}  in the
$\Lambda$CDM cosmology with
WMAP parameters \citep{wmap7}. The  reference spectrum is the blackbody CMB spectrum
  at $z=10^7$. At $x<0.01$ the spectrum is dominated by
  bremsstrahlung and double Compton processes that destroy the condensing low-frequency photons. The effect of bremsstrahlung compared to
  {Comptonization} becomes stronger with decreasing redshift and can be felt
  at successively higher  frequencies. 

At $z\gtrsim 5\times 10^4$ the spectral distortions in
  the $x\gtrsim 0.01$ region can relax to a Bose-Einstein spectrum due to
  high value of parameter $y$. Contributions from lower redshifts are
  described by the analytic solution of Eq.~\eqref{szap}.
  The distortions created at $z<10^4$ are just {the} negative of the usual
  {SZ} effect and would be
completely {overwhelmed} by the {SZ} effect from reionization at
$z\sim 10$ which is expected to be  $\YSZ =(\kB  \Te/\me c^2) \tau_{ri}\sim
10^{-6}\times 0.1=10^{-7}$, where $\tau_{ri}\sim 0.1$ is the optical depth
to the last scattering surface due to reionization and $\Te\sim 10^4\rm{K}$
is the average electron temperature during reionization. The distortions created by energy loss at $z>10^5$ can, however, be
destroyed  only by energy injection at $z>10^5$ since at lower
redshift {Comptonization} cannot create $\mu$-type distortions.

\section{Energy release from dissipation of acoustic waves
  during radiation domination}
In the standard cosmology the {main} source of energy injection at high redshifts
is the dissipation of
acoustic waves in the baryon-photon plasma \citep{sz1970} due to Silk
damping \citep{silk}. \citet{sz1970b} first proposed  using the resulting
{spectral} distortions to measure the spectral index and power in the small-scale
fluctuations, which themselves do not survive. Later during
recombination the second-order Doppler effect due to non-zero electron/baryon velocity in the CMB
rest frame also becomes important, but at that time only {$y$-type distortions similar to the SZ effect arise}, and $\mu$ distortions cannot be created.
We can {estimate} the amount of $\mu$ distortions created by dissipation of acoustic waves
following \cite{hss94}
by calculating how the total power in the density fluctuations changes with
time because  of photon diffusion.
The \emph{comoving} energy {density} in acoustic waves in the photon-baryon plasma (neglecting {the} baryon
energy density)  is given by $Q=\rho_{\gamma}c_s^2 \langle
\delta_{\gamma}(\mathbf{x})^2\rangle$, where $\rho_{\gamma}$ is the
comoving photon
energy density, $c_s^2\sim 1/3$ is sound speed squared, $\delta_{\gamma}$
is the photon density perturbation at position $\mathbf{x}$, and angular
brackets denote the ensemble average:
\begin{align}
\langle \delta_{\gamma}(\mathbf{x})^2\rangle= \int \frac{\id^3k}{(2\pi)^3}P_{\gamma}(k),
\end{align}
where $P_{\gamma}(k)=\Delta_{\gamma}^2(k)P_{\gamma}^i(k)$, and
$P_{\gamma}^i(k)$ is the initial power spectrum. It was shown by
\citet{mukhanov} that primordial fluctuations from inflation can have a
spectrum deviating from  the scale-invariant Harrison-Zeldovich
spectrum, with spectral index $n_s<1$. In the cyclic ekpyrotic models
$n_s>1$  is also 
possible \citep{turok}. WMAP gives the constraint
on curvature perturbation in comoving gauge, $\zeta$. This is related to the
 gravitational perturbation $\psi$ in the radiation era  (assuming
 neutrinos are free streaming) by the relation
 $\psi=\zeta/(2/5R_{\nu}+1.5)$, where
 $R_{\nu}=\rho_{\nu}/(\rho_{\gamma}+\rho_{\nu})\approx 0.4$, $\rho_{\nu}$ is
 the neutrino energy density and
 $\delta_{\gamma}^i=-2\psi$ \citep{mabert95}. {Thus
 $P_{\gamma}^i=4/(2/5R_{\nu}+1.5)^2P_{\zeta}=1.45P_{\zeta}$ and
 $P_{\zeta}=(A_{\zeta}2\pi^2/k^3)(k/k_0)^{\nS -1+\frac{1}{2} r (\ln k/k_0)}$, $k_0=0.002{\,{\rm Mpc^{-1}}}$,
 $A_{\zeta}=2.4\times 10^{-9}$ (\citep{wmap7,spt}), $r=\rm{d} \nS /\rm{d}\ln k$ is the
 running of the index.} The transfer function
 for modes well inside the horizon before recombination
is given by
\begin{align}
\Delta_{\gamma}\approx 3\cos (kr_s)e^{-k^2/\kD ^2}\label{tf}
\end{align}
and the diffusion {scale by} \citep{kaiser,weinberg}
\begin{align}
\frac{1}{\kD ^2}&=\int_z^{\infty}\id z
\frac{c(1+z)}{6H(1+R)\Ne \sigT}\left(\frac{R^2}{1+R}+\frac{16}{15}\right)\\
\frac{\id (1/\kD ^2)}{\id z}&=-\frac{c(1+z)}{6H(1+R)\Ne \sigT}\left(\frac{R^2}{1+R}+\frac{16}{15}\right){,}
\end{align}
where $R\equiv3\rho_b/4\rho_{\gamma}$, $\rho_b$ is the baryon energy
density.

{Replacing $\cos^2(kr_s)$ with} its average value over an oscillation of $1/2$, we get
the energy release per unit redshift

\begin{align}
\frac{\id Q/\id z}{\rho_{\gamma}}&=\frac{-1}{3}\int
\frac{\id^3k}{(2\pi)^3}P_{\gamma}^i(k)\frac{\id\Delta_{\gamma}^2}{\id z}\nonumber\\
&=3\int
\frac{\id^3k}{(2\pi)^3}P_{\gamma}^i(k)k^2e^{-2k^2/\kD ^2}\frac{\id(1/\kD ^2)}{\id z}\nonumber\\
&=\frac{4.3 A_{\zeta}}{k_0^{\nS -1}}\frac{\id(1/\kD ^2)}{\id z}\int
 \id k \, k^{\nS }e^{-2k^2/\kD ^2}\nonumber\\
&=\frac{4.3 A_{\zeta}}{k_0^{\nS -1}}\frac{\id(1/\kD ^2)}{\id z}2^{-(3+\nS )/2}\kD ^{\nS +1}\Gamma\left(\frac{n+1}{2}\right){.}
\end{align}
{For a running index, the above integration must be done numerically.}
We are interested in radiation-dominated epoch where $\mu$ distortions are
generated, and we have $H(z)=H_0\Omega_r^{1/2}(1+z)^2$ and
$\Ne (z)=(n_{H0}+2n_{He0})(1+z)^3\equiv n_{e0}(1+z)^3${. In addition} we have
$\kD =A_D^{-1/2}(1+z)^{3/2}$ and $\id(1/\kD ^2)/\id z=-3A_D(1+z)^{-4}$, and the above
equation simplifies to
\begin{align}
\frac{\id Q/\id z}{\rho_{\gamma}}&=-\frac{13 A_{\zeta}}{k_0^{\nS -1}}2^{-(3+\nS )/2}\Gamma\left(\frac{\nS +1}{2}\right)A_D^{(1-\nS )/2}(1+z)^{(3\nS -5)/2}\nonumber
\end{align}
where
\begin{align}
A_D&=\frac{8 c}{135 H_0\Omega_r^{1/2}n_{e0}\sigT}=5.92\times
10^{10}\rm{Mpc}^2{.}
\end{align}
The redshift and $n_s$ dependence obtained above matches that of \citet{hss94}.
The diffusion scale $\kD =46\rm{Mpc}^{-1}$ at $z=5\times 10^4$ and
$\kD =10^4\rm{Mpc}^{-1}$ at $z=2\times 10^6$. Thus we are probing the primordial
fluctuations on very small scales that are not accessible in any other way.

\begin{figure}
\includegraphics{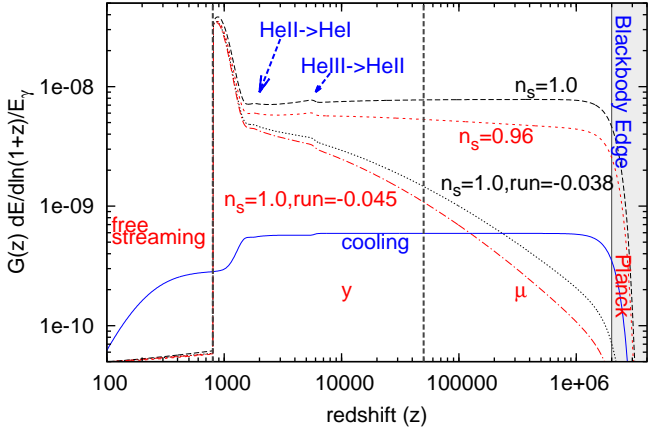}
\caption{\label{energyfig}Sketch of fractional rate of energy release due to Silk damping and free
  streaming for different initial power spectra.  Also shown for comparison is the rate of energy loss due
  to adiabatic cooling of baryonic matter.  }
\end{figure}

Thus for {a} Harrison-Zeldovich spectrum with  $\nS $=1 
\begin{align}
\frac{\id Q/\id z}{\rho_{\gamma}}&=\frac{-7.8\times 10^{-9}}{1+z}\nonumber\\
\frac{Q_{z=2\times 10^6}^{z=5\times 10^4}}{\rho_{\gamma}}&=2.9\times 10^{-8}{.}
\end{align}
We note the surprising fact that this redshift dependence is exactly
the same as for the energy
losses due to adiabatic cooling Eq.~\eqref{energy}, {as already pointed out by \citet{cs2011}}.
For the currently {preferred} value of $\nS =0.96$ we have
\begin{align}
\frac{\id Q/\id z}{\rho_{\gamma}}&=\frac{-1\times 10^{-8}}{(1+z)^{1.06}}\nonumber\\
\frac{Q_{z=2\times 10^6}^{z=5\times 10^4}}{\rho_{\gamma}}&=1.8\times 10^{-8}{.}
\end{align}
Figure \ref{energyfig} compares the rate of energy release multiplied by the
blackbody visibility function,
$G(z)(1+z)\id Q/\id z/E_{\gamma}$ for different
initial power spectra with the energy losses due to the adiabatic cooling
of baryons, $(1+z){S_{\rm Compton}}/E_{\gamma}$ obtained by solving
Eq.~\eqref{tee} for the standard recombination history  \citep{sss2000}  
calculated  using the effective multilevel
approach and taking  recent corrections into account \citep{ah2011,ct2011}. The spike due to increase in Silk damping during recombination
is clearly noticeable\footnote{\label{foot}The actual energy injected in
    the recombination peak would be much  less because  of the transition to free streaming.}. The smaller spikes due to HeII and HeI
  are also noticeable. Recombination of each species leads to a decrease in
  the number of particles,  hence to adiabatic losses. But the photon diffusion
  length, and the associated energy release,  increases owing to a decrease in
  the electron fraction. The transition from radiation to matter domination
  results in unnoticeable change in the slope of the curves.  Even after recombination photons continue
to mix on horizon scales because of free streaming, while the energy
losses from Comptonization drop sharply as a result of depletion of
electrons. We switch to the free-streaming solution described in appendix
\ref{appa} when diffusion length $2\pi/\kD$ becomes equal to the comoving
particle horizon. In the blackbody
  photosphere region, the blackbody visibility function $G(z)$ makes the curves drop
  sharply. Global energy release and $y$-type distortions during the free
  streaming epoch after recombination are due to the superposition of blackbody
  spectra on horizon scales. The energy injected into $y$-type distortion
  during recombination in the peak  ($800<z<1500$) is $\Delta
  E/E\sim 10^{-8}$ for $n_s=0.96$ (see also footnote \ref{foot}). The $y$ distortion due to mixing of
blackbodies during free streaming  after recombination and up to reionization
, $20<z<800$, calculated using the result in the
appendix \ref{appa}, is $\Delta E/E\sim 10^{-10}$. These are, however, difficult
to separate from much the larger  thermal and non-linear Doppler  $y$ distortion from reionization of
$Y_{SZ}\sim 10^{-7}$.

\begin{figure}
\includegraphics{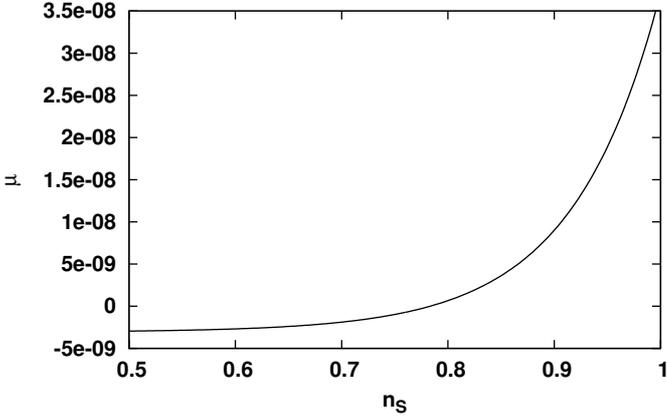}
\caption{\label{ns} $\mu=1.4\Delta E/E$ as a function of spectral index
  $\nS$ without running. }
\end{figure}

\begin{figure}
\includegraphics{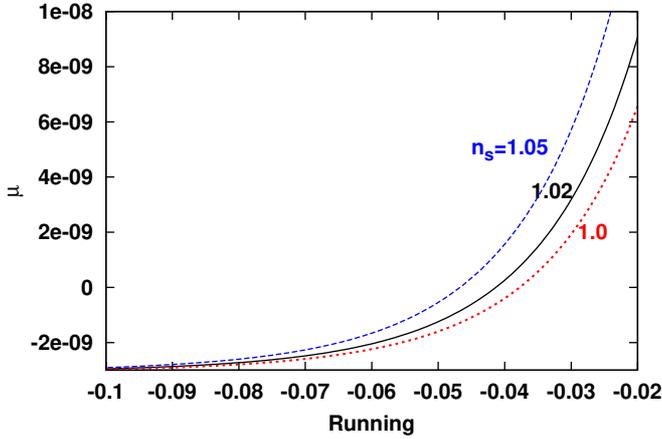}
\caption{\label{nrun}$\mu=1.4\Delta E/E$ as a function of running of the spectral index
  $r$ for three values of spectral index $\nS$.}
\end{figure}

Energy releases for different power spectra, including the ones with running
spectral index, are summarized in {Tables \ref{tab1} and \ref{tab2}.} 
The energy losses from adiabatic cooling of baryons during the same time
period is $\sim 2.2 \times 10^{-9}$. {This equals the energy release, for
example, 
for a  spectral index of $\nS =1.0,\rm{d} \nS/\rm{d}\ln
k=-0.038$ for a running spectrum. The net $\mu$ distortion as a function of spectral index
  $\nS$ without running is shown in Fig.
  \ref{ns} and for $\nS=1.0,1.02,1.05$ as a function of the running of the
  index in Fig. \ref{nrun}. For low values of $\nS$ and high 
  negative values of running, Bose-Einstein condensation dominates, and net distortion
  approaches a constant value of $-3\times 10^{-9}$.}
A detection or non-detection at a sensitivity of $10^{-9}$ of $\mu$-type distortion is thus a very sensitive probe of primordial
power spectrum  on small scales.  Given the importance of
the effective heating rate for the amplitude of the net distortion, it is
important to carry out a more careful and refined calculation of the
problem, which we present elsewhere {\citep{cks2012}.}

{
Over the years a number of ground- and balloon-based experiments that aimed at measuring the absolute CMB
brightness have been conducted before COBE \citep{exp1,exp2}
and after it \citep{exp3,exp4,exp5,exp6,exp7,exp8,tris1,arcade}. The best limits on CMB $\mu$-type spectral distortions
from these experiments are at the level of $10^{-4}-10^{-5}$ \citep{cobe,arcade2,tris2}.
Achieving a sensitivity of $10^{-9}$ would require subtracting 
foregrounds due to synchrotron emission, free free emission, dust emission,
and spinning dust emission at the same precision level. The proposed experiment
Pixie \citep{pixie} aims to achieve this goal by using 400 effective
channels between frequencies of 30GHz to 6THz. Their simulations indicate
that an accuracy of 1nK in foreground subtraction is achievable. There is, however, some uncertainty
in our understanding of the foregrounds and possible systematics as
indicated by an observed excess signal at 3GHz by ARCADE \citep{arcade}
experiment, which is not completely explained
by the current galactic and extragalactic emission models; nevertheless achievement
of a  goal of $10^{-9}$ in sensitivity, although challenging, seems possible in
the near future.
}

\begin{table}
\begin{tabular}{|c|c|}
\hline
$ \nS $ & $ \Delta E/E$ \\
\hline
\color{blue}$ 1.07$ &\color{blue} $ 6.8\times 10^{-8}$\\
\color{blue}$ 1.04$ &\color{blue} $ 4.7\times 10^{-8}$\\
\color{blue}$ 1.0$ &\color{blue} $ 2.9\times 10^{-8}$\\
\color{blue}$ 0.96$ & \color{blue}$ 1.8\times 10^{-8}$\\
\color{blue}$ 0.92$ & \color{blue}$ 1.1\times 10^{-8}$\color{black}\\ 
\hline
\color{red} BEC & \color{red}$ -2.2\times 10^{-9}$\color{black}\\ 
\hline
\end{tabular}
\caption{\label{tab1}Energy injection in $\mu$ distortions during $5\times 10^4<z<2\times 10^6$
  for different {initial power spectra without running} compared with energy losses due
to Bose-Einstein condensation.} 
\end{table}
\begin{table}
\begin{tabular}{|c|c|c|}
\hline
$ \nS $ &  $ \rm{d} \nS /\rm{d}\ln k$ & $ \Delta E/E$\\
\hline
\color{blue}$ 1.07$ &-0.05&\color{red} $ 2.2\times 10^{-9}$\\
\color{blue}$ 1.07$ &-0.035& \color{blue}$ 5.7\times 10^{-9}$\\
\color{blue}$ 1.07$ &-0.02& \color{blue}$ 1.6\times 10^{-8}$\\
\color{blue}$ 1.04$ &-0.05&\color{red} $ 1.6\times 10^{-9}$\\
\color{blue}$ 1.04$ &-0.035& \color{blue}$ 4.1\times 10^{-9}$\\
\color{blue}$ 1.04$ &-0.02& \color{blue}$ 1.1\times 10^{-8}$\\
\color{blue}$ 1.0$ &-0.05&\color{red} $ 1.1\times 10^{-9}$\\
\color{blue}$ 1.0$ &-0.038& \color{red}$ 2.2\times 10^{-9}$\\
\color{blue}$ 1.0$ &-0.035& \color{blue}$ 2.6\times 10^{-9}$\\
\color{blue}$ 1.0$ &-0.02& \color{blue}$ 6.9\times 10^{-9}$\\
\hline
\color{red} BEC &-& \color{red}$ -2.2\times 10^{-9}$\color{black}\\ 
\hline
\end{tabular}
\caption{\label{tab2}Energy injection in $\mu$ distortions during $5\times 10^4<z<2\times 10^6$
  for different {initial power spectra with running} compared with energy losses due
to Bose-Einstein condensation. Energy injection values $\le$ BEC value in magnitude are shown in red.} 
\end{table}
\section{Conclusions}
 Comptonization of CMB photons with slightly cooler electrons results in
 spectral distortions that  then evolve (in the absence
of any other physical process) towards the 
Bose-Einstein condensate solution, which is the
equilibrium
solution  and   consists of a blackbody spectrum with zero chemical
potential, with the extra photons accumulating at zero frequencies \citep{is1975b,llstats}. This is
because the radiation  has more photons than can be accommodated
in a Bose-Einstein spectrum with non-negative chemical potential at the electron temperature. 
However, in reality, bremsstrahlung and double Compton scattering  destroy extra photons
at low frequencies. Thus in practice we  have the high-frequency
spectrum evolving towards {a} blackbody with the extra photons slowly moving
down in frequency to be eaten up by bremsstrahlung and double Compton scattering. This is  what is
seen in the numerical solution in Fig. \ref{becfig}.

The difficulty of conserving photon number is the reason  Bose-Einstein condensation (or accumulation of photons at low frequencies) is
not observed in astrophysical systems. Bose-Einstein condensation of
photons has only been achieved in the laboratory  very recently by
\citet{ksvw2010}. In the real Universe we also have inevitable sources of energy
 injection into the CMB, such as the  SZ effect after reionization, which creates distortions of similar
magnitude but with opposite signs {canceling} and overwhelming  this
effect.  The situation is much more interesting for $\mu$ distortions, which
can be created only at high redshifts and thus have no foregrounds at
lower redshifts. In standard cosmology, positive $\mu$ distortions are created by
dissipation of acoustic waves {on} small scales due to photon diffusion. {For
the currently allowed values of  cosmological parameters, there is a
possibility that the positive $\mu$
distortions from photon diffusion can almost exactly cancel the negative $\mu$
distortions from Bose-Einstein condensation of CMB, leading to a net
distortion that is much smaller. Bose-Einstein condensation can even
dominate, leading to a net distortion $\mu \sim -3\times 10^{-9}$ to which  Silk damping
 contributes, but only as a small perturbation.} Nevertheless, we must
emphasize that  the photon number
conserving Comptonization, along with differences in adiabatic indices of
radiation and matter, creates a unique system in the early Universe in which
photons can begin to evolve towards a Bose-Einstein
condensate. It is a remarkable coincidence that a completely unrelated
physical process,  diffusion damping of primordial perturbations, can 
produce distortions of almost exactly the same or greater magnitude, leading to the
suppression of this unique effect in astrophysics. 
We thus arrive at an
intriguing conclusion: even a null result, non-detection of $\mu$-type
distortion at a sensitivity of $10^{-9}$, rather than just placing a upper
limit, actually gives a quantitative measure of the
primordial small-scale power spectrum. The importance of improving the
experimental sensitivity to reach this critical value cannot be overemphasized.
\begin{appendix}
\section{Erasure of perturbations due to free streaming}\label{appa}
We can derive the solution for erasure if perturbations on horizon scales
due to free streaming as follows.
We start with the Boltzmann hierarchy, ignoring quadrupole and higher order
moments. We define multipole ($\ell$) moments of the temperature anisotropy
$\Delta T/T\equiv \Theta =
\sum_{\ell}(-i)^{\ell}(2\ell+1)\mathcal{P}_{\ell}(\hat{k}.\hat{n})\Theta_{\ell}$,
where $\mathcal{P}_{\ell}$ is the Legendre polynomial, and
$\hat{k}.\hat{n}$  the angle between photon momentum and comoving
wavenumber. In the limit of zero Thomson optical depth we get
\begin{align}
\frac{\rm{d}\Theta_0}{\rm{d}\eta}+k\Theta_1=0\nonumber\\
\frac{\rm{d}\Theta_1}{\rm{d}\eta}-\frac{k}{3}\Theta_0=\frac{k\psi}{3},\nonumber
\end{align}
where $\eta=\rm{d}t/a$ is the conformal time.
Using the fact that in the matter-dominated era the potential is suppresses
by a factor of $9/10$ compared to primordial value on  large scales, we
have $\Theta_0=-2\psi/3$ \citep[see for example][]{dod}
and looking for a solution with time dependence of type $\Theta_{\ell}\propto e^{i\int d\eta \omega}$
we get  $\omega=ik/\sqrt{6}$
Thus free streaming damps the perturbations by a factor of $e^{-\int d\eta
  k/\sqrt{6}}=e^{-0.4\int d\eta k}$.
Finally we want to mention that this solution is very approximate, and in
principle we should  also take higher $\ell$ modes into account. For example, on going
up to $\ell=3$ we get for the damping factor  $e^{-0.3\int d\eta k}$.
Also for free streaming we have super position of blackbodies, and the $y$
distortion is given by \citep{Chluba2004}
$\YSZ=1/2<\Theta^2>=1/32<\delta_{\gamma}^2>$, and so the equivalent energy
release is 
$\Delta E/E=4\YSZ=1/8<\delta_{\gamma}^2>$.
\end{appendix}
\begin{acknowledgements}
We would like to thank Matias Zaldarriaga for important remarks on the
manuscript. We would also like to thank Yacine Ali-Ha\"{i}moud for
careful reading and comments on the manuscript.
\end{acknowledgements}
\bibliographystyle{aa}
\bibliography{bec}
\end{document}